# Brain Tumor Recurrence vs. Radiation Necrosis Classification and Patient Survivability Prediction

M. S. Sadique, W. Farzana, A. Temtam, E. Lappinen, A. Vossough, K. M. Iftekharuddin

*Abstract*— GBM (Glioblastoma multiforme) is the most aggressive type of brain tumor in adults that has a short survival rate even after aggressive treatment with surgery and radiation therapy. The changes on magnetic resonance imaging (MRI) for patients with GBM after radiotherapy are indicative of either radiation-induced necrosis (RN) or recurrent brain tumor (rBT). Screening for rBT and RN at an early stage is crucial for facilitating faster treatment and better outcomes for the patients. Differentiating rBT from RN is challenging as both may present with similar radiological and clinical characteristics on MRI. Moreover, learning-based rBT versus RN classification using MRI may suffer from class imbalance due to lack of patient data. While synthetic data generation using generative models has shown promise to address class imbalance, the underlying data representation may be different in synthetic or augmented data. This study proposes computational modeling with statistically rigorous repeated random sub-sampling to balance the subset sample size for rBT and RN classification. The proposed pipeline includes multiresolution radiomic feature (MRF) extraction followed by feature selection with statistical significance testing ($p<0.05$). The five-fold cross validation results show the proposed model with MRF features classifies rBT from RN with an area under the curve (AUC) of 0.892±0.055. Moreover, considering the dependence between survival time and censor time (where patients are not followed up until death), we demonstrate the feasibility of using MRF radiomic features as a non-invasive biomarker to identify patients who are at higher risk of recurrence or radiation necrosis. The cross-validated results show that the MRF model provides the best overall performance with an AUC of 0.77±0.032. Comparison with state-of-the-art methods suggest the proposed methods are effective in RN versus rBT classification and patient survivability prediction.

This work was supported in part by NIH/NIBIB grant under award number R01EB020683, and by the National Science Foundation Grant 1828593. (*Corresponding author:* Dr. Khan Iftekharuddin)

M. S. Sadique, W. Farzana, A. Temtam and, K. M. Iftekharuddin are with Vision Lab and Department of Electrical and Computer Engineering, Old Dominion University, Norfolk, VA 23529 USA (e-mail: msadi002@odu.edu; wfarz001@odu.edu; atemt001@odu.edu; kiftekha@odu.edu).

E. Lappinen is with Radiation Oncology and Biophysics, Eastern Virginia Medical School, Norfolk, VA 23507 USA (e-mail: LappinEK@evms.edu)

A. Vossough is with Department of Radiology at Children's Hospital of Philadelphia, 3401 Civic Center Boulevard, Philadelphia, PA 19104 USA (e-mail: VOSSOUGH@chop.edu)

## I. INTRODUCTION

GLIOBLASTOMA multiforme (GBM) is the most common and aggressive primary brain tumor in adults. These tumors are highly invasive and difficult to treat due to their ability to spread quickly throughout the brain, making them a significant challenge for medical professionals. Patients with GBM have a poor prognosis even with the standard treatment of maximal safe surgical resection followed by concurrent chemoradiotherapy (CCRT) with temozolomide (TMZ) or adjuvant TMZ. The median overall survival time is only 14-16 months, and the 2-year survival rate is only 26-33% [1, 2]. It is imperative to make a prompt and accurate diagnosis of postoperative progression in order to select the most appropriate treatment plans and thereby maximize the likelihood of patient survival.

Radiation necrosis, the most common complication in gliomas within 2 years of treatment, is a common result of this combination therapy of CCRT with TMZ. Radiation necrosis typically occurs during the peak time for glioma recurrence [3, 4]. MRI and PET scans can diagnose radiation necrosis, but they can also confuse it with a recurrent tumor. In the clinic, recurrence of a glioma is distinguished from necrosis by follow-up observations, biopsy, and resection [5] . Screening for recurrent brain tumors and radiation necrosis at an early stage is crucial for facilitating faster treatment, better outcomes, lower morbidity and mortality, enhanced quality of life, and tracking treatment efficacy.

The development of more advanced and more efficient imaging techniques has contributed significantly to an increase in the accuracy of disease diagnosis. Although the mechanisms and methods that are currently in use provide valuable information on the phenotypes of cancer, the specifics of many of the cases are not yet fully resolved. Many difficulties remain in applying data-driven disease detection techniques in practice. In recent years, the medical field has made widespread use of computer-aided diagnostic techniques for disease diagnosis and treatment planning. However, there is frequently an inequitable distribution of data in medical databases. As a result, class imbalance appears to be a prevalent phenomenon in the datasets used for disease classification [6-9].

Radiomic features can be extracted from medical images and can provide valuable information about the underlying tissue characteristics. Radiomics has the potential to improve diagnosis, treatment planning, and patient outcomes in a variety of diseases[10-12]. Finding a fast and noninvasive way to differentiate radiation necrosis from tumor recurrence is important because the treatment protocols for these two conditions are very different [13-17]. Lesion recurrence or necrosis can be classified noninvasively using radiomics features [17-19]. However, these radiomics studies use relatively small sample sizes, which limits the generalizability of their results to larger populations. Moreover, these models have several challenges including bias due to underrepresentation of minority classes, overfitting, and the inability to generalize the model's performance.

We address the problem of class imbalance by combining methods from three different levels: feature selections, random sampling, and ensemble learning, in order to develop a robust ensemble model that outperforms state-of-the-art competitive base and ensemble classifiers. Identifying patients who are at low risk of recurrence or radiation necrosis could also potentially lead to more personalized and less aggressive treatment plans, improving the quality of life for these patients while still ensuring their long-term survival.

In addition to accurately predicting patient survival outcomes, Copula-based survival analysis models have been shown to improve accuracy over conventional techniques, especially in situations involving non-proportional covariate effects or non-linear dependence between time-to-event and covariates. In contrast to Cox modeling, which makes a strong independence assumption, Copula-based modeling is appropriate if there is a possibility of dependence between survival time and censoring time [20-25]. We consider radiomic feature selection procedures for survival data with dependent censoring. Dependent censoring occurs when the probability of censoring depends on the survival time and/or the value of a covariate. Conventional survival analysis methods may introduce bias, so Copula-based feature selection procedures can be very useful in these circumstances. Through radiomic analysis, we used a Copula graphic estimator multivariate feature selection method that considers the dependence structure of the censoring mechanism and shows its superior performance compared to existing methods. The proposed method is then applied to aggressive brain tumor datasets to demonstrate its effectiveness in identifying important features associated with survival outcomes. According to the findings, radiomic features may be helpful in predicting the likelihood of recurrence or radiation necrosis in patients, which could help guide treatment decisions and lead to better patient outcomes. The proposed method has been implemented in an R Compound Cox package.

The goal of this work is to identify efficient and reliable methods to deal with the challenges posed by class imbalance due to the scarce availability of radiation necrosis patient data and to develop a model for the classification of glioma recurrence versus radiation necrosis. We demonstrated the potential of MRI multiresolution texture features for prediction of rBT versus RN. Further, this study aims to ascertain the efficacy of the MRF features for patient survivability prediction. The findings of this study can help clinicians to better identify patients who require more aggressive treatment and those who may benefit from less invasive approaches. This can ultimately lead to improved patient outcomes and reduced healthcare costs.

## II. BACKGROUND STUDY

In this part, we discuss class imbalance learning and dependent censoring in classification and survival analysis. Class imbalance learning refers to the problem of uneven distribution of classes in a dataset, which can lead to biased model performance. Dependent censoring, on the other hand, occurs when the censoring mechanism is related to the underlying survival time, making it challenging to accurately estimate survival probabilities.

### A. Class Imbalance Learning

Class imbalance in machine learning has attracted a lot of attention, particularly in the medical field where accurate classification of minority classes is crucial. An imbalance between the classes may occur when one class possesses a significantly lower number of samples than the other. Most classification methods suffer from subpar results on unbalanced datasets [26-31] because they are modeled on the assumption of class-balance. This is especially true when the imbalance ratio is high. Several machine learning methods for classification of imbalanced datasets of various disease phenotypes have been developed [32-35]. Rodriguez-Almeida et al., report training four different supervised machine learning classifiers using synthetically generated datasets alone and in combination with the real data to ensure data integrity and consistent classification performance [36].

One of the key machine learning methods is ensemble learning. Combining resampling with ensemble learning has the potential to reduce the degree of overfitting in the minority class while still reserving all cases for the majority. However, ensemble learning has drawbacks, such as lengthy training times and high computational complexity. Recent studies have focused on a few imbalanced learning schemes, including UnderBagging, SMOTEBagging, RUSBoost, and SMOTEBoost. In RUSBoost, training subsets are generated serially via a sampling method, a classifier instance is trained, and both the classifier's performance and the difficulty of each training pattern are evaluated [37-39]. Synthetically generated oversampling can easily lead to overfitting issues, especially for highly imbalanced datasets. While undersampling may result in the loss of information from a majority class, this can be effectively mitigated by combining ensemble learning and data distribution design [27, 28, 36].

In many real-world situations, the class imbalance cannot be solved by using just one strategy. Instead of using only one unbalanced classifier to perform risk assessment of tumor failure in head and neck cancer using radiomics, Vallieres, et al. [33, 34] propose a different ensemble of multiple balanced classifiers using imbalance adjustment in the training process. This is done as an alternative to using only one unbalanced classifier. The generic ensemble imbalance learning (EIL) framework MESA proposed by Liu et al. [40] uses data-driven learning to optimize imbalanced classification via a meta-sampler to incorporate adaptive under-sampling into iterative ensemble training. The overall distribution of classification errors is entered at each iteration and used to train a new base classifier, which is then merged with the ensemble to produce a new state. This will enhance the effectiveness of generalization.

Underrepresented groups suffer from the effects of class bias in the form of subpar model performance. Our study used a method called "repeated random sampling" to achieve a more even distribution of classes by selecting random samples from the original data set in a standardized manner. Using this strategy can help lessen the effects of bias and broaden the applicability of the findings.

### B. Dependent Censoring

When patient follow-up ends before the patient's expected time of death, this is called censorship. Censoring time is an important concept in survival analysis, as it allows us to account for incomplete data and estimate the probability of an event occurring beyond the observed time period. It is typically used in longitudinal studies and clinical trials to ensure that the data collected is accurate and representative of the population being studied.

Dependent censoring has an effect that may help with prediction accuracy. Logic suggests that dependent censoring due to patient dropout is informative about survival predictability because it is tied to the patients' health status. By contrasting a prognostic index with a censoring index on a graph, Siannis et al. [41] propose a graphical diagnostic method. If there is a positive correlation between the censoring index and the prognostic index, then a high censoring index indicates a high mortality risk. The purpose of survival analysis is to provide an estimate of the hazard function, which is the instantaneous risk of an event occurring at a given time. The survival function, which measures the likelihood of surviving past a specified time, is related to the hazard function. In order to estimate the hazard function and predict survival probabilities, Copula models [21-23, 25] offer a flexible framework for modeling the dependence between the survival time and censoring time.

### III. MATERIALS AND METHODOLOGY

#### A. Study Population and MRI Images

In this study, the patient data set is collected from the Cancer Genome Atlas (TCGA) in the Genomic Data Commons (GDC) Data Portal [42]; the Cancer Imaging Archive (TCIA) [42-46]; and in an Institutional Review Board-approved (Eastern Virginia Medical School (EVMS) IRB# 20-6-NH-0130), joint study from collaborating institutions: EVMS, Sentara Healthcare and Old Dominion University (Vision Lab).

A total of 158 patients who had surgery first and then went on to receive radiation treatment for their glioma were included in this study. During the period of follow-up, there were 15 patients who showed signs of radiation necrosis and 143 patients who showed tumor recurrence. Following treatment, every patient went through a series of scans that included T1-weighted precontrast (T1), T1-weighted postcontrast (T1C), T2-weighted (T2), and fluid-attenuated inversion recovery (FLAIR).

In our study, out of a total of 158 patients, 44 have died, 43 have been lost to imaging follow-up (11 are dead and 32 alive), and 114 are alive. We considered days to death (i.e., censorship = 1) or loss to follow up (i.e., censorship = 0). In Fig. 1 a flowchart depicting the number of patient's included for each analysis is shown. This chart illustrates the patient selection procedure and ensures inclusivity in the data collection and analysis.

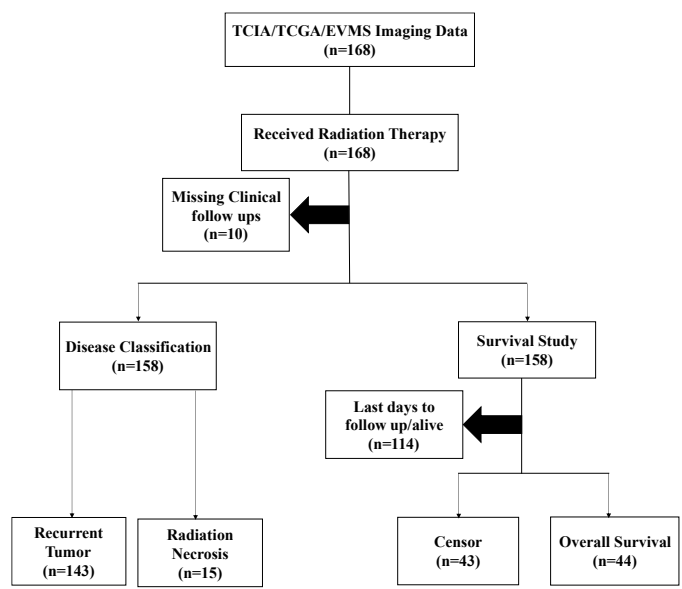

Fig. 1. Flowchart of patients included and excluded for each analysis.

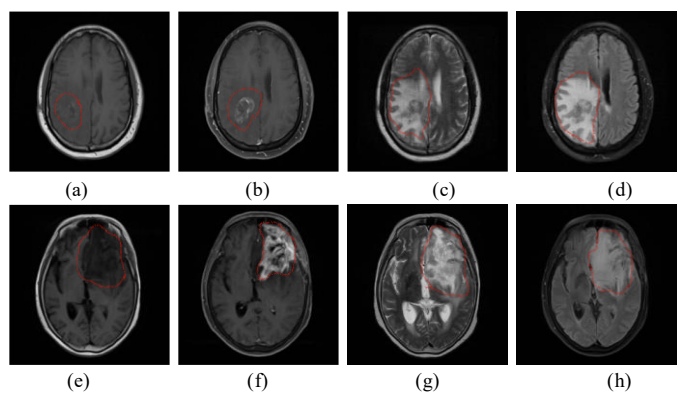

Fig. 2. MRI of two patients with glioma (T1, T1C, T2, FLAIR modalities) (a-d) Radiation Necrosis Case after 8 months follow up and (e-h) Recurrent Case after 6 months follow up

The following criteria are used to select patients for inclusion. All glioma patients' recurrence or necrosis after radiotherapy is confirmed by imaging and clinical follow-up; all MRI images (T1, T1C, T2, and FLAIR images) used must be confirmed at a follow-up after radiotherapy; and glioma patients without pathologic diagnosis are excluded from the analysis. A neuroradiologist's ability to determine whether a patient is experiencing recurrence or radiation necrosis depends on the length of time between scans. Patients with recurrent glioma who did not receive radiotherapy, were followed for less than 6 months, or had not undergone all four modalities of magnetic resonance imaging were not included. The axial plane is utilized for the acquisition of each MRI scan. An example of each of the four different modalities of MRI images showing glioma recurrence and necrosis is presented in Fig. 2.

## B. Image Preprocessing

We resample the MRI images to $1mm^3$ resolution and perform skull stripping to remove the skull and non-brain tissues from the MRI images after registering T1, T2, and FLAIR images to T1C images using the linear registration function in FSL5.0.9 (http://fsl.fmrib.ox.ac.uk) [47]. For each patient, we isolate the area of the tumor using our 3D deep learning model [48, 49]. A neuroradiologist verifies the segmented mask to ensure its accuracy. The whole tumor, the tumor core, and the enhanced tumor region are all included in the segmentation used to extract the features.

Our method is made up of three components: the selection of features, repeated sub-sampling, and the proposed ensemble algorithms. The subsequent sections will introduce each of the three parts. The overall working flow diagram for the proposed method is shown in Fig. 3. The significance test and decision tree classification methods are used in conjunction with the data-level method that is based on feature selection to select the characteristics that are the most important.

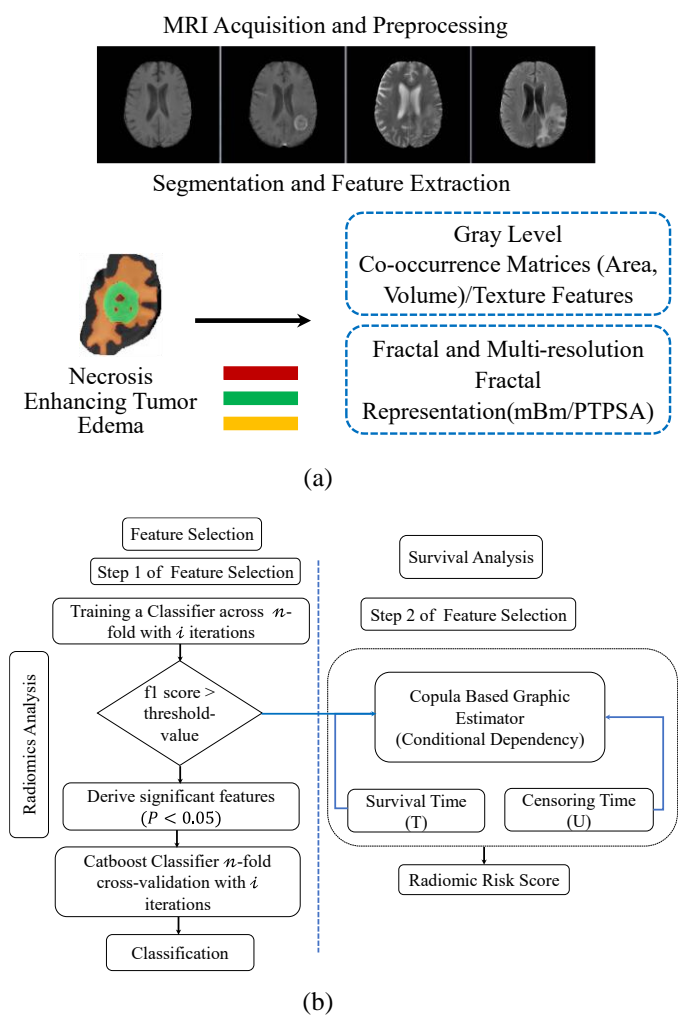

Fig. 3. Overall pipeline for the proposed method (a) Segmentation and Feature extraction (b) Radiomics workflow for classification and survivability prediction

## C. Feature Extraction

Texture features are extracted from the whole tumor (WT) volumes (including edema, enhance tumor, and necrosis) using MRI techniques such as T1, T1C, T2, and FLAIR. These techniques produce histogram-based statistics and matrix-based features such as the co-occurrence matrix, neighborhood gray tone difference matrix, and the Size Zone Matrix [12, 50]. The histogram-based statistics from the different tumor sub-regions including enhance tumor (ET), edema (ED), and necrosis (NCR) are also used to extract features from the T1, T1C, T2, and FLAIR MRI sequences (ET, ED, NCR) [48, 49]. Volumes of the different regions (ET, ED, NCR) are also described by extracting a number of volumetric features [51-53]. Finally, the tumor volume is analyzed along three dimensions (x, y, and z) to determine nine spatial features (area, centroid, perimeter, major axis length, minor axis length, eccentricity, orientation, solidity, and extent). To further facilitate the extraction of the texture features, we employ a fractal and multi-resolution fractal representation of the tumor volume. Fractal and multi-resolution fractal representations of tumor volume across the four modalities include the piecewise triangular prism surface area (PTPSA), multifractional Brownian motion (mBm), and the Holder exponent (HE) [32, 50-57].

## D. Feature Selection

It is essential to keep in mind that the selection of features involves making a compromise between the number of features employed and the level of precision achieved by the model. Too few features may lead to a model that is too simple and has poor accuracy, while too many features may lead to a model that is too complex and overfits the data. Both outcomes are possible depending on the number of features included in the model. As a result, it is essential to carefully evaluate the outcomes of feature selection to ascertain the optimum number of features to use in the context of the particular data set and problem that is being investigated.

In our study feature selection is accomplished with the help of a decision tree classifier by analyzing the weight given to each characteristic during the process of determining the target variable. One method for determining which features are the most relevant and important in terms of making accurate predictions of the target variable is to use a decision tree classifier and select features based on their F1 scores. It is common practice to employ the F1 score when evaluating the efficiency of binary classification models because it provides a measurement of the equilibrium between precision and recall. One strategy for selecting features based on the F1 score when using a decision tree classifier is to iteratively remove the least important feature, retrain the model, and calculate the F1 score after each removal. Another strategy is to use an ensemble of multiple decision trees. This process can be repeated as many times as necessary until all features have been examined. The feature that has the effect of lowering the F1 score by the least amount is considered as the most important feature, while the feature that has the effect of lowering the F1 score by the most amount is considered as the least important feature [58-60]. Then, a threshold (0.85 for MRF and 0.80 for NFRF) can be set based on the scores to select the top k most important features, where k is a number that is user-defined.

### E. Repeated Random Sub-sampling

Random sub-sampling may be used repeatedly to deal with highly unbalanced data sets [38, 40, 61, 62]. It is crucial to pay attention to the class distribution when dealing with medical data because most classification algorithms assume that the class distribution in the data set is uniform. This method divides the data set into majority and in majority instances, from which the training and testing data sets are generated. Using this repeated subsampling method, the training data is partitioned into sub-samples, with each sub-sample containing an equal number of instances from each class, except for the last sub-sample (in some cases). This sampling approach is used to ensure that each sub-sample is representative of the overall population. Every subsample is used as a test case for fitting the classification model, and the result is obtained by a majority of the subsamples.

---

**Algorithm 1** Sample ( $\mathcal{D}_\mathcal{T}$ )

---

Require: ( $\mathcal{D}_\mathcal{T}$ )
1. Initialization: derive majority sample $\mathcal{N}_1$ and minority sample $\mathcal{N}_O$ from $\mathcal{D}_\mathcal{T}$
2. Assign sample subset $(x_i, y_i)$ in $\mathcal{N}_1$
3. Sample majority class $\mathcal{N'}_1$ from $\mathcal{N}_1$ w.r.t. equal subsets, where $|\mathcal{N'}_1| = \mathcal{N}_O$
4. **return** balanced subset $\mathcal{D'}_\mathcal{T} = \mathcal{N'}_1 \cup \mathcal{N}_O$

---

In this work, we use the following repeated random sub-sampling approach for randomly dividing the dataset into training and testing sets multiple times, allowing for a more robust evaluation of the model's performance. The whole dataset samples $\mathcal{D}_\mathcal{T}$ are divided into two separate data sets, $\mathcal{N}_1$ majority data samples and $\mathcal{N}_0$ in minority data samples, where $\mathcal{D}_\mathcal{T} = \mathcal{N}_0 + \mathcal{N}_1$. The majority training samples $\mathcal{N}_1$ are uniformly distributed across all training data subsamples, while the minority samples $\mathcal{N}_O$ are sampled without replacement. The data sampling algorithm is shown in Algorithm 1.

---

**Algorithm 2** $n$-fold training with $i$ iterations

---

Require: $\mathcal{D}_\mathcal{T}, C$
1. for iteration $k$=0 to $i$-1 do
2. for fold $nj$=1 to n-1 do
3. for each balanced subset $\mathcal{D'}_\mathcal{T} = \mathcal{N'}_1 \cup \mathcal{N}_O$
4. train $\mathcal{C}_1(\mathcal{X})$ classifier across $n$ -fold with $i$ iterations
5. **return** ensemble $\mathcal{C}_\text{E}(\mathcal{X}) = \sum_{k}^{i-1} \sum_{j}^{n-1} \mathcal{C}_j(\mathcal{X})$

---

### F. Classification Model

Algorithm 2 shows the training procedure of a classifier for $n$-fold cross validation with $i$ iterations. Using gradient boosting decision trees, CatBoost [63, 64] is an ensemble method for machine learning. The ensemble method involves combining the predictions of multiple decision trees that have been trained independently on separate subsets of the training data. CatBoost employs gradient boosting, which means it seeks to enhance prediction accuracy by mitigating mistakes made by earlier decision trees in an ensemble. The ensemble's final prediction is a weighted average of the individual decision trees' predictions, with higher weights given to the decision trees that performed the best.

### G. Survival Analysis Modeling

Survival analysis often uses time-to-death event times. The best quantitative indicator of cancer patients' well-being is overall survival (OS). Survival analysis is made more difficult by censorship. When a study stops taking new patients or when a patient drops out, it omits their survival data. When examining survival data, medical researchers and statisticians typically make the assumption that censoring mechanisms are unrelated to the event of interest. Statistical independence between events and censoring times is taken for granted by survival analysis methods like the Kaplan-Meier estimator and Cox regression.

Censoring may bias statistical analysis if symptoms worsen, and patients drop out or withdraw. Informative dropout is one of many censoring factors. Dependent censoring occurs when any event mechanism censors the event time of interest. Under the independent censoring assumption, survival analysis methods are unbiased. This study attempts to identify imaging biomarkers that can help to better understand the heterogeneity of brain tumors and improve the accuracy of prognostic models for survival analysis. The Copula method allows us to identify the most relevant features for predicting survival outcomes in censored patients.

We consider two random variables; survival time T refers to the length of time of an individual or group survives from a specific event or diagnosis and censoring time U is the point at which data collection stops, often due to the end of a study or loss to follow-up. We assume that survival time T and censoring time U are conditionally independent given a feature $x_j$ for all j =1,…, p. Rather than assuming complete independence, it is more realistic to assume independence only under certain conditions.

Let's consider T is the amount of time spent alive, U is the amount of time spent being censored, and x is a vector of censoring time T and U are assumed to be conditionally independent if and only if there is no correlation between them and any of the components of x. This assumption is typically made for Cox regression models. Also, let the marginal survival functions given x be denoted by $S_T(t|x) = P_r(T > t|x)$ and $S_U(u|x) = P_r(U > u|x)$ where the probability is given as,

$$P_r(T > t, U > u|\mathbb{x}) = C_\theta \{S_T(t|\mathbb{x}), S_U(u|\mathbb{x})\}; \quad (1)$$

where $C_\theta$ is a Copula function [23, 25] and $\theta$ is a parameter characterizing the degree of dependence between *T* and *U*. Here, a bivariate survival function is taken into consideration. According to this model, the dependency between *T* and *U* is characterized by the variable $C_\theta$. The marginal distribution for each variable and the Copula function that represents the dependence structure between them make up the two main parts of the Copula model. The Copula function illustrates the interdependence of the variables represented by the marginal distributions of survival time and censoring time.

The conditional independence assumption is invalid as T and U are not independent when only $x_j$ is known. The correlation between two variables, T and U, can be evaluated using Kendall's $\tau$ (s), which is defined as,

$$\tau = P_r\{(T_2 - T_1)(U_2 - U_1) > 0 | \mathbb{x}\} - P_r\{(T_2 - T_1)(U_2 - U_1) < 0 | \mathbb{x}\}; \quad (2)$$

where $(T_1, U_1)$ and $(T_2, U_2)$ are values derived from the model (1), respectively.

### H. Statistical Significance Analysis

Comparative analyses of radiomics features with recurrent GBM and patients with RN are performed using independent t-tests for continuously distributed normally and utilizing Mann–Whitney U-tests for continuously distributed variables that do not follow a normal distribution. P-value < 0.05 is selected as the appropriate threshold of statistical significance.

### I. Classification Performance Evaluation

Based on the confusion matrix, the performance of the classification is assessed using rates of accuracy (ACC), area under the curve (AUC), positive predicted value (PPV), and false positive rate (FPR). For the comparison with our proposed learning model with imbalanced dataset, we further apply two different synthetic data generation methods. To balance the datasets, two widely used methods are applied: SMOTE (Synthetic Minority Over-sampling Technique) [28] and ADASYN (Adaptive Synthetic Sampling) [27]. SMOTE generates synthetic samples by interpolating between existing minority class samples, while ADASYN focuses on generating more synthetic samples in regions where the density of minority class samples is low. Both methods aim to improve the performance of machine learning models on imbalanced datasets. Moreover, for the classification task, we compare our experiments with other works in the literature for classification of recurrent brain tumor from radiation necrosis. The 5-fold cross-validation method is used to assess each of the proposed models.

## IV. RESULTS AND DISCUSSION

This section discusses two representative experiments in this study as follows.

### A. Experiment 1: Classification of Recurrent Brain Tumor from Radiation Necrosis in GBMs

In this experiment, the relative importance of the features is calculated first using corresponding F1 score, and the results are ranked. Statistical analysis shows that 8 (P < 0.05) out of the 27 and 7 (P < 0.05) out of 30 are the most important features from each of the MRF and NFRF methods, respectively. The results of our classification experiments are summarized in Table I. We implement $n$ = 5-fold cross validation with $i$ =1000 iterations. Radiomics analysis is used to determine whether a glioma is necrotic or has recurred. We choose two overesampling methods from the imbalanced-learn Python package [68] to evaluate the efficacy of the proposed framework for addressing class imbalance in the patient data. We assess the effectiveness of the proposed framework by calculating metrics like area under the curve (AUC), accuracy, and (PPV).

Table I summarizes the experimental results from this work and comparison with other works in the literature for classification of recurrent brain tumor from radiation necrosis that use resampling methods to address limited sampling issues [65-67]. The results show that repeated random subsampling strategy, learned from a small number of training instances, outperforms other data resampling methods. Consequently, even though these resampling methods generate and use a huge number of synthetic samples for training, they rely heavily on relations between minority classes and, hence, worsen the classification performance. Note that due to the differences in MRI modalities, methodologies, and datasets, it may not be possible to draw direct comparisons between these studies and our proposed method in this study.

TABLE I
COMPARISON OF CLASSIFICATION PERFORMANCE OF THE PROPOSED FRAMEWORK WITH OTHER METHODS FOR RECURRENT TUMOR VS RADIATION NECROSIS CLASSIFICATION

| Model Configurations | Method | Dataset size | Area Under Curve (AUC) | Accuracy (%) | Positive Predicted Value (PPV) | False Positive Rate (FPR) |
|---|---|---|---|---|---|---|
| **Proposed MRF Model** | **Repeated Random Subsampling** | **RN=143, rBT=15** | **0.89±0.055** | **80.88±0.061** | **0.80±0.064** | **0.15±0.07** |
| **Proposed NFRF Model** | **Repeated Random Subsampling** | **RN=143, rBT=15** | **0.86±0.061** | **77.43±0.064** | **0.77±0.064** | **0.16±0.08** |
| **Proposed SMOTE-MRF** | **SMOTE** | **RN=143, rBT=15** | **0.835** | **91.00** | **0.73** | **-** |
| **Proposed ADASYN-MRF** | **ADASYN** | **RN=143, rBT=15** | **0.85** | **80.00** | **0.63** | **-** |
| Random Forest (Xuguang, et al.[65]) | SMOTE | RN=37, rBT(TP)=98 | 0.71 | - | - | - |
| Random Forest (Prateek, et al. [66]) | - | RN=22, rBT=33 | 0.79 | - | - | - |
| SMOTE (Park, *et al.* [67]) | SMOTE | RN=18, rBT=23 | 0.80 | 78.00 | - | - |

The optimal feature combinations for the prediction estimation of multiresolution and nonfractal radiomics models are shown in Fig. 4.

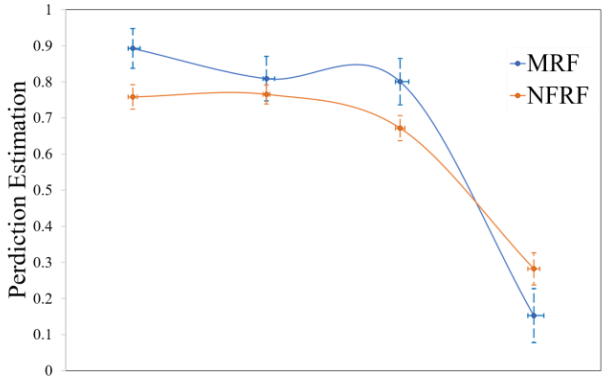

Fig. 4. Multi-texture model prediction estimation (AUC, ACC, PPV, FPR) assessment using the following feature sets: (MRF) and (NFRF).

It shows that in the MRF model, the AUC is 0.89±0.055, the Positive Predicted Value (PPV) is 0.80±0.064, and the False Positive Rate (FPR) is 0.15±0.074 better when compared to the NFRF prediction estimation.

### B. Experiment 2: Survival Prediction and Patient Outcome

In our study out of the total of 158 patients, 44 have passed away, 43 have been lost to follow-up, and 114 have neither. Therefore, a total of 87 (44 + 43) patients are used for the survival analysis and prediction. The standard practice for selecting important features for survival analysis in the absence of censoring survival data or independent censoring [21] is to use a univariate Cox proportional hazard model. In Censorship or lack of follow-up meant that those patients are excluded from our analysis.

From the total number of clinical patient data collected for this study we know that 43 patients who were lost to follow-up are alive. We propose using the Copula method for feature selection in the censoring scenario. We begin by extracting radiomic features and determining the dependency parameter Θ (via the survival data). The p-vector of features for these 87 patients is written as $X_i = (x_{i1}, x_{i2},..., x_ip)$. Then, the significant features (p-value < 0.05) are used to predict the probability of survival outcomes and to make a binary prediction of whether the patient is expired or dead (denoted as 1) or not expired or dead (denoted as 0).

The strategy proposed here aims to consider the impact of dependent censoring on biomarkers that may potentially affect the marginal survival. The experiments and statistical significance analysis demonstrate that the selected features may be used to develop a predictive model for patient outcomes. A plot of the cross-validated c-index utilizing the training samples is presented in Fig. 5. The c-index is at its highest point when the association parameter is equal at α=18 which corresponds to Kendall's tau=0.90. In dependent censoring, the c-index is physically significant because it quantifies a model's discrimination ability relative to the risk of the event of interest. Better discrimination between subjects with different risks of experiencing the event of interest is indicated by a higher c-index.

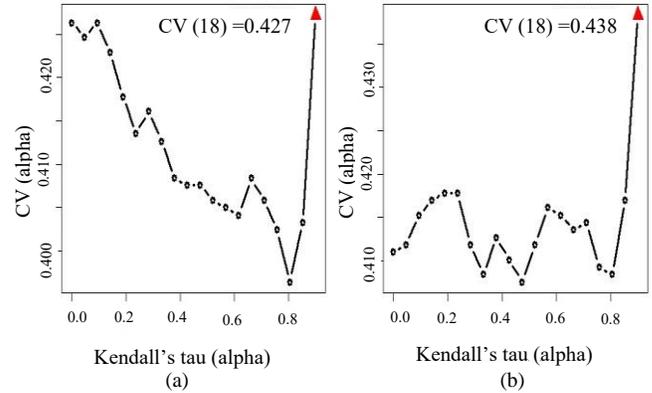

Fig. 5. The cross validated c-index for the MRF and NFRF models. The maximum value of the cross-validated c-index can be found at α = 18 which corresponds to Kendall's tau = 0.90. (a) c-index = 0.427 for MRF model, (b) c-index = 0.438 for NFRF model.

We plot the survival curves for the good (or poor) prognosis (Fig. 6) groups differentiated by low (or high) values of the PIs in order to verify the predictive power of the most significant features on overall survival. These groups are distinguished by the severity of their disease. This is done so that we can assess how well each category is doing in terms of how long they live. To correct for the inherent bias in the Kaplan-Meier survival curve when independent censoring is in play, we use the Copula-graphic estimator of the survival curves as modified by the Clayton Copula at α=18 (Kendall's tau=90). We perform the survival prediction analysis for the MRF and NFRF models using the different numbers of significant feature sets (P-value < 0.05).

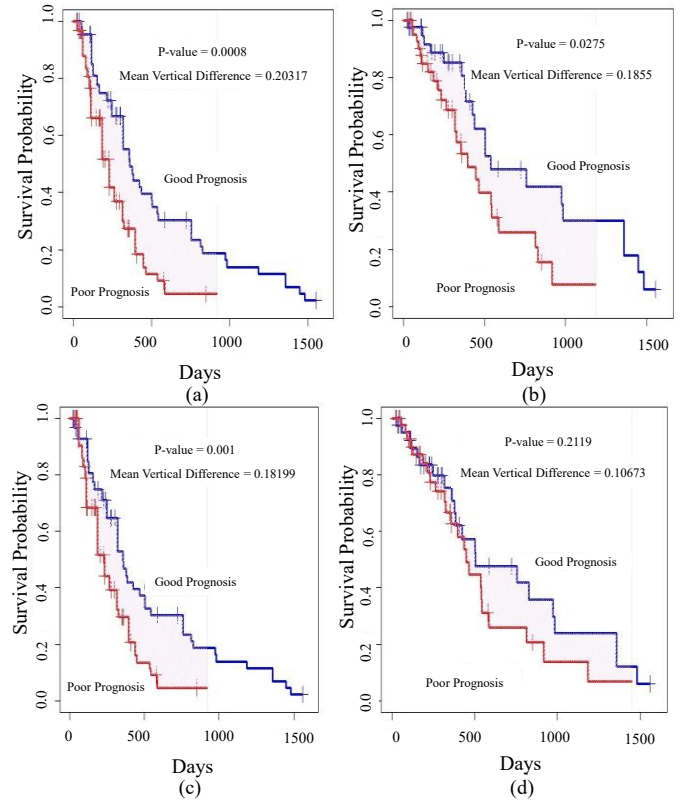

Fig. 6. The marginal survival curves for the group with a good (or poor) prognosis determined by the most significant features. MRF model: :(a) Dependent censoring (multivariate), (b) Independent censoring (univariate). NFRF model: (c) Dependent censoring (multivariate), (d) Independent censoring (univariate).

The results for different feature combination are summarized in Table II. The average vertical difference between the survival curves over the study period is used to determine the degree of separation between the curves, and the corresponding P-value is calculated using the permutation test. Smaller P-values correspond to a better separation of the patients with a good prognosis from those with a poor prognosis. In most cases, the proposed method produces significant separation between the patients with a good prognosis and those with a poor prognosis (P-value < 0.05) while using multivariate analyses (dependent censoring) (Table II).

TABLE II
MEAN VERTICAL DISSIMILARITY BETWEEN GOOD AND POOR PROGNOSIS GROUPS IN SURVIVAL CURVES. THE P-VALUE THAT WAS DETERMINED THROUGH A PERMUTATION TEST BY RANDOMLY DIVIDING ALL OF THE SAMPLES INTO SAMPLES OF THE SAME SIZE.

(A) MRF MODEL

| Feature Numbers | Dependent Censoring Mean Diff.(P-value) | Independent (Univariate) Mean Diff. (P-value) |
|---|---|---|
| 3 | 0.14979 (0.0051) | 0.19639(0.0183) |
| 5 | 0.11327 (0.0353) | 0.215 (0.0101) |
| 7 | 0.14233 (0.0084) | 0.215 (0.0093) |
| 10 | 0.16046 (0.0046) | 0.1414 (0.0955) |
| 16 | 0.20317(0.0008) | 0.1855 (0.0275) |

(B) NERF MODEL

| Feature Numbers | Dependent Censoring Mean Diff.(P-value) | Independent (Univariate) Mean Diff. (P-value) |
|---|---|---|
| 3 | 0.14978 (0.0051) | 0.1766(0.0325) |
| 5 | 0.11327 (0.0353) | 0.1709 (0.0413) |
| 7 | 0.14233 (0.0084) | 0.08285 (0.3303) |
| 10 | 0.16046 (0.0046) | 0.10634 (0.2107) |
| 12 | 0.18199(0.001) | 0.10673 (0.2199) |

According to our findings, the MRF model is effective at determining which features are most important for making accurate prognostications about patient outcomes. The coefficient value obtained from radiomic features in the dependent censoring setting is statistically significant. Table III summarize the 16 important features with their coefficient value. The MRF model suggests that these 16 significant features are very useful for predicting the outcome of the patient. In this case, the Copula feature selection (multivariate) method provides an extremely clear separation between the good and poor prognosis patients (P-value =0.008 as shown in Fig. 6). In case of NFRF model, the Copula feature selection (multivariate) method leads to a better separation of the good and poor prognoses (P-value = 0.001; Fig. 6 (c) and (d)) compared to that for univariate selection (P-value = 0.2199). Figure 6 shows that, in contrast to the case of univariate selection, the MRF and NFRF perform significantly better at distinguishing between good and poor prognoses when subjected to dependent censoring. Hence, the list of features selected by the multivariate Copula method is consistently more predictive than those selected by univariate selection. This suggests that the proposed methods are more effective in identifying potential radiomic biomarkers that are associated with good and poor prognoses. Therefore, they may be useful in developing more accurate treatment plans to significantly improve the diagnosis and management of recurrent brain tumors and radiation necrosis.

TABLE III
THE 16 MOST STRONGLY ASSOCIATED MRF FEATURES BASED ON MULTIVARIATE METHODS: DEPENDENT CENSORING. THE FEATURES ARE ORDERED ACCORDING TO THE P-VALUES.

| Features | Coefficient | P- Value |
|---|---|---|
| T1C_GTSDM_Difference_Entropy | -1.47 | 0.0030 |
| T1_37GTSDM_InverseDifference | 1.18 | 0..0058 |
| T1_8Histogram_Entropy | -1.41 | 0.0080 |
| T1_ptpsa_GLZSM_GrayLevelNonUniformity | 1.10 | 0.0166 |
| T1_GTSDM_AngularSecondMoment | 1.63 | 0.0170 |
| T1C_GTSDM_Cluster_Prominence | -1.44 | 0.0174 |
| Fl_GQV_GLZSM_LargeZoneLowGrayEmphasis | 1.34 | 0.0175 |
| T1_GLZSM_ZoneSizePercentage | -1.06 | 0.0192 |
| T1_GLZSM_LargeZoneSizeEmphasis | 1.05 | 0.0333 |
| T1C_37GTSDM_Entropy | -0.99 | 0.0333 |
| T2_mBm_GTSDM_Information_Correlation1 | -1.39 | 0.0348 |
| T2_38GTSDM_Difference_Entropy | -1.06 | 0.0378 |
| T1C_mBm_GTSDM_InverseDifferencemoment | 1.12 | 0.0383 |
| T1_37GTSDM_Entropy | -0.91 | 0.0387 |
| Fl_GQV_GLZSM_LargeZoneHighGrayEmphasis | 1.01 | 0.0440 |
| T1C_GLZSM_LargeZoneSizeEmphasis | 0.91 | 0.0450 |

In survival analysis, a prognostic index (PI) is a numerical score that is used to predict the outcome or prognosis of a patient based on multiple prognostic factors or variables. Prognostic index is calculated from the selected radiomic features. The PI values are commonly used in clinical practice to aid decision-making and to stratify patients into risk groups. They can also be useful in designing treatment planning by identifying patients who are more likely to benefit from a treatment.

In this study (Fig. 7), we find the number of patients with a good or poor prognosis using the prognostic indices. PI is high for poor prognosis/high risk groups and PI is low for good prognosis/low risk groups. This information can help us identify patients who may require more intensive treatment or monitoring and those who may have a better chance of recovery. It also highlights the importance of early detection and intervention in improving patient outcomes.

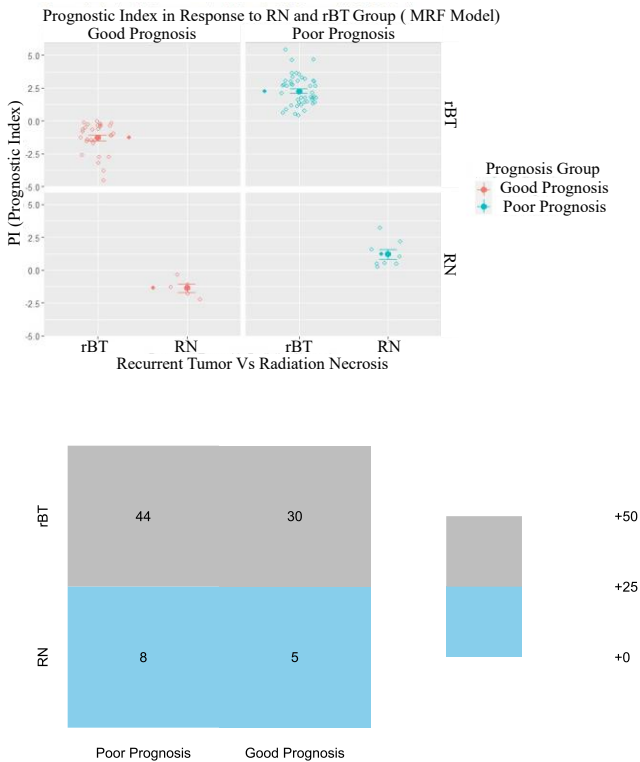

Fig. 7. Two-way plot for prognostic index in response to good or poor prognosis

Figure 7 shows the total number of patients in each category. It is evident from the plot that most patients have a high prognostic index value which are associated with worse survival outcomes. This means that they are more likely to experience disease recurrence, radiation necrosis, or death compared to patients with a low score. In this case, a high prognostic index value could be used to identify patients who may require for targeted interventions to improve their outcomes.

Table IV provides a summary of the findings for binary classification in rBT and RN patients (dead or alive) using different combinations of significant feature sets associated with the prognostic assessment of rBT and RN. When predicting rBT and RN patient survival classification based on MRF results in a highest area under the curve (AUC) of 0.77±0.032 (PPV= 0682±0.030), whereas classification based on nonfractal radiomic features results in an area under the curve (AUC) of 0.758±0.033 (PPV = 0.672±0.034) (Table IV). Fig. 8 shows the predictive properties of the MRF and NFRF models for the binary prediction of patient outcomes using different combination of significant feature sets. The results in Table IV suggest that MRF features have a higher predictive power than nonfractal radiomic features for predicting patient survival classification in rBT and RN patients. In conclusion, we demonstrate that multiresolution (MRF) radiomics features model provide the important prognostic information for the risk assessment of rBT and RN.

TABLE IV
COMPARISON OF DIFFERENT CONFIGURATIONS OF THE MODEL BASED ON THE MEAN TEST PERFORMANCE ACROSS $n$ = 5-FOLD WITH $i$ = 1000 ITERATIONS

| Number Of features | Models | Area Under Curve (AUC) | Accuracy (%) | Positive Predicted Value (PPV) | False Positive Rate (FPR) |
|---|---|---|---|---|---|
| Top 3 features | MRF | 0.73±0.035 | 72.61±0.027 | 0.62±0.033 | 0.34±0.046 |
| | NFRF | 0.67±0.036 | 70.77±0.026 | 0.60±0.032 | 0.36±0.043 |
| Top 5 features | MRF | **0.77±0.032** | **76.98±0.025** | **0.68±0.030** | **0.35±0.041** |
| | NFRF | 0.73±0.035 | 71.99±0.026 | 0.61±0.034 | 0.35±0.048 |
| Top 7 features | MRF | 0.77±0.032 | 76.91±0.025 | 068±0.031 | 0.27±0.041 |
| | NFRF | 0.76±0.032 | 75.99±0.025 | 0.66±0.033 | 0.29±0.042 |
| Top 10 features | MRF | 0.75±0.033 | 76.21±0.025 | 0.66±0.033 | 0.28±0.042 |
| | NFRF | **0.75±0.033** | **76.54±0.026** | **0.67±0.034** | **0.28±0.044** |

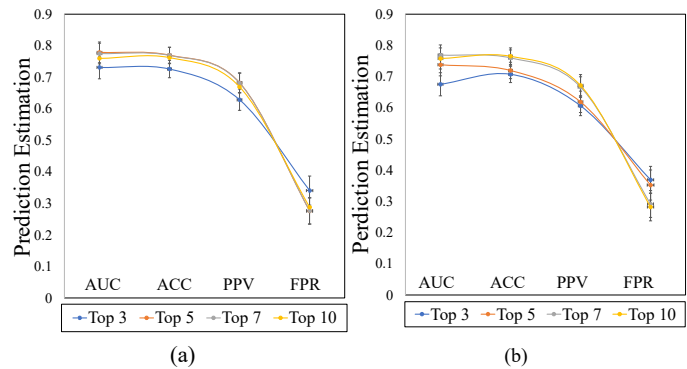

Fig. 8. Inspection of predictive properties of multivariable survival models constructed from 2 feature sets: (a) MRF, (b) NFRF.

## V. CONCLUSION

In conclusion, this work shows that the proposed MRF radiomics feature may be used to classify RN vs. rBT for GBM patients. We further present an efficient ensemble learning model to handle the class imbalance problem. Our findings demonstrate that the optimal radiomic variables combined with repeated random sampling may have a favorable effect on the prognostic evaluation of recurrent brain tumors and radiation necrosis with limited patient data as shown in Table I. These findings suggest that this type of analysis using MRI may be an additional useful tool for the diagnosis and prognostication of recurrent brain tumors, potentially resulting in more patient-specific treatment options.

We demonstrate our proposed ensemble model combining feature selection and repeated sub-sampling can effectively predict survivability. In our study, the MRF-based model outperforms the nonfractal NFRF-based model for distinguishing recurrent brain tumor from radiation necrosis in GBMs, with areas under the curves, PPV, and FPR of 0.892±0.055, 0.800±0.064, and 0.152±0.0740, respectively. This work further obtains comparison of the prediction performance of the proposed learning models with other similar methods in the literature for differentiating rBT from RN (Table I). The comparison results show that the proposed learning model outperforms the other models in terms of AUC for differentiating recurrent brain tumor from radiation necrosis.

We further assess potential biomarker selection strategies for survival data with dependent censoring. Identifying the most crucial radiomic features requires us to use both univariate (independent censoring) and multivariate (dependent censoring) feature selection methods. Our results and simulations show that the proposed methods are more useful for finding potential radiomic biomarkers associated with good and poor prognoses when dependent censoring is present. For future work, the proposed models need to be generated and validated using larger patient cohorts.